\documentclass[envcountsect,envcountsame,runningheads]{llncs}
\usepackage{tikz-cd}
\usepackage{algorithm}
\usepackage{algpseudocode}
\usepackage{amsmath,amssymb}
\usepackage{graphicx}
\usepackage[colorlinks=true, allcolors=black]{hyperref}
\usepackage{algorithm}
\usepackage{algpseudocode}
\usepackage{xspace}
\newcommand{\brev}{\color{black}}
\newcommand{\erev}{\color{black}}

\def\AlgebraicSolvers{\href{https://github.com/AlgebraicGeometricModeling/AlgebraicSolvers.jl}{AlgebraicSolvers.jl}\xspace}

\usepackage{documenter}

\usepackage{graphicx} \usepackage{amsfonts,amsmath,amssymb}

\usepackage{amsthm}

\usepackage{cleveref}
\usepackage{tikz-cd}
\usepackage{xspace}
\usepackage{kbordermatrix}

\usepackage{graphics}

\theoremstyle{plain}

\newcommand{\eb}{\ensuremath{\mathbf{e}}}

\newcommand{\Ac}{\ensuremath{\mathcal{A}}}

\newcommand{\Dc}{\ensuremath{\mathcal{D}}}

\renewcommand{\epsilon}{\ensuremath{\varepsilon}}
\newcommand{\CC}{\ensuremath{\mathbb{C}}} 
 
\newcommand{\kk}{\ensuremath{\mathbb{K}}} 
\newcommand{\okk}{\ensuremath{\overline{\mathbb{K}}}}

\newcommand{\QQ}{\ensuremath{\mathbb{Q}}}  
\newcommand{\NN}{\ensuremath{\mathbb{N}}}

\newcommand{\scp}[1]{\langle #1 \rangle}
\newcommand{\vspan}[1]{\langle #1 \rangle}
\newcommand{\set}[1]{\{ #1 \}}
\newcommand{\dual}[1]{ #1^{*}}
\newcommand{\cl}[1]{\mathcal{#1}}

\newcommand{\bm}[1]{\mathbf{#1}}
\newcommand{\bms}[1]{\boldsymbol{#1}}

\def\kkrng{\kk[\bm x]}

\def\:{\!:\!}

\def\eval{\eb}

\newcommand{\PolExp}{\cl Pol\cl{E}xp}
\newcommand{\invsyst}[1]{{\langle\langle{#1}\rangle\rangle}}

\def\C1{{\cl C}(1)\,}

\usepackage[english]{babel}

\usepackage[letterpaper,top=2.5cm,bottom=2.5cm,left=3.5cm,right=3.5cm,marginparwidth=1.95cm]{geometry}

\begin{document}

\title{Eigensolvers for polynomial roots and tensor decomposition}
\author{Enrica Barrilli \and Bernard Mourrain}
\institute{Centre Inria at Universit{\'e} C{\^o}te d'Azur}

\maketitle

\begin{abstract}
Computing eigenvalues and eigenvectors is at the heart of the solution of many non-linear problems.
For instance, finding the roots of polynomial systems reduces to computing joint eigenvectors of operators of multiplication.
Similarly, tensor decomposition can be performed via the joint diagonalization of submatrices of the Catalecticant of the tensor.
We describe and illustrate symbolic-numeric methods
for computing the solutions of these algebraic problems from the computation of joint eigenvectors of commuting operators, and for analysing their multiplicity structure, as well as their implementation in the package \AlgebraicSolvers
\end{abstract}

\section{Introduction}

Nonlinear problems are ubiquitous in scientific computing. However, solving them efficiently often relies on the use of fundamental tools from linear algebra, such as solving linear systems, or computing eigenvalues or eigenvectors.

In this article, we aim to illustrate this "motto", focusing on the solutions of algebraic problems. A typical algebraic problem is finding all the roots of a set of polynomial equations. It has applications in many domains, e.g. in robotics, computer vision, chemistry, signal processing, polynomial optimization, \dots (see e.g. \cite{Sommese2005,Elkadi2007,Kukelova2008,cox2020applications,Lasserre2009,Dumitrescu2017}). 

Various methods exist to find the roots of polynomial equations.
The most popular are probably local iterative methods, such as Newton-Raphson method \cite{Dedieu2015}, which start from an initial point and usually converge to a fixed point.
Such methods are very efficient in practice to find a single solution, but the convergence is not guaranteed and depends highly on the initialization. Moreover, they find one solution and have difficulties to find all the solutions of a system.

Homotopy solvers constitute another family of solvers \cite{Sommese2005}. They exploit local iterative methods in a controlled way, by deforming a system with known solutions into the system to be solved, tracking the paths of the solutions by local methods. They are efficient, but can suffer from the proximity of singularities. 

We are interested in Algebraic solvers, which exploit the properties of the quotient algebra $\cl A=R/I$ of the ring of polynomials $R$ by the ideal $I$ generated by the system of equations to solve \cite{CLO}.
They make it possible to calculate all the roots of a zero-dimensional system and to analyze their multiplicities. 

The \AlgebraicSolvers package is dedicated to this family of methods. It is based on the concept of 
Truncated Normal Form (TNF) \cite{Mourrain1999,Telen2018,Mourrain2021,mourrain:hal-05135952}, which provides a uniform approach to handling this family of solvers.
Tools for computing TNFs based on the computation of Gröbner bases are available (see Section \ref{sec:mult roots}). This computation has been optimised over several decades \cite{Faugre2010,Eder2017,msolve,demin2023groebner} to achieve a high level of performance, but it applies to polynomial equations with exact coefficients (for example, coefficients in $\QQ$).
However the TNF can be derived using numerical linear algebra (with floating-point arithmetic), once the Gr\"obner basis has been computed for any order. 
Border bases, developed to address the problem of numerical instability in Gröbner bases 
\cite{Mourrain1999,Mourrain2005,Kreuzer2008},
 can be used in the same way to compute a TNF.

Resultant matrix constructions, which have also been devised to solve specific classes of polynomial systems (e.g. generic polynomials with a given support) \cite{Emiris1999,cox2020applications} can also be used to deduce a TNF. The TNF is computed from the co-kernel of the resultant matrix, again using numerical linear algebra.
In the package \AlgebraicSolvers, the solvers are parametrized by classes, which specify how to compute the TNF (see Section \ref{sec:examples}).

TNFs can also be used in Tensor Decomposition Problems 
(see \cite{Iarrobino1999,brachat2010symmetric,bernardi2013general} and Section \ref{sec:tensor dec}) or in Polynomial Optimization Problems \cite{Lasserre2009} for recovering the optimizers. 
They can be computed directly from the Hankel or Catalecticant matrix associated to the tensor or from the optimal moment matrix in Polynomial Optimization Problems (see \cite{Henrion2005,mourrain:hal-05135952}).

Truncated Normal Forms provide an effective description of the multiplicative structure of the quotient algebra $\cl A$ and, in particular, of the operators of multiplication by elements in $\cl A$.
The solutions of the polynomial systems or the nodes in a tensor decomposition problem can then be recovered by eigencomputation \cite{Auzinger1988,moller1995multivariate,Mourrain1998,Stetter2004,Elkadi2007}. We recall these properties in Section \ref{sec:artinian} and illustrate the TNF approach in Section \ref{sec:examples}.
For simple roots, their coordinates are computed by join diagonalization of the matrices of multiplication by the variables.
Direct eigenvector computation (using the function \texttt{LinearAlgebra.eigen}) of a random combination of the matrices or iterative numerical methods (see \cite{vanderhoeven:hal-01579079,cardoso1996jacobi,bunse1993numerical}) are implemented. In the presence of multiple roots, a Schur factorization is performed (using \texttt{LinearAlgebra.schur}) and the coordinates of the roots are recovered using traces of the diagonal blocks of the multiplication matrices (see \cite{Corless1997,barrilli:hal-05339397}).
To analyze further the multiplicity of a root, inverse systems  can also be computed, using the function \texttt{AlgebraicSolvers.invsys} based on linear algebra tools (see \cite{macaulay1916algebraic,Mourrain1998,hauenstein:hal-01250388,mantzaflaris:hal-03079910,mantzaflaris:hal-02478768}).

These methods are based on the use of duality properties on polynomial rings and on the representation of linear functionals as formal power series (see \texttt{AlgebraicSolvers.Series}).  

\section{The structure of an Artinian algebra}\label{sec:artinian}
An \emph{Artinian} algebra $\cl A$ is a commutative $\kk$-algebra of finite dimension as a $\kk$ vector space: $\dim_\kk \cl A < \infty$. As it is finite dimensional, $\cl A$ is also a finitely generated algebra. Suppose that the Artinian algebra $\cl A$ is generated by $a_{1}, \ldots, a_{n}$ as an algebra. Then we have the exact sequence:
$$
\begin{array}{rcl}
    0 \longrightarrow  I \longrightarrow  \kk[x_{1},\ldots,x_{n}] &\longrightarrow & \cl A \longrightarrow  0\\
    x_{i} & \longmapsto & a_{i} \\
\end{array}
$$
where $I$ is a zero-dimensional ideal \textcolor{black}{in the ring $R$ of polynomials in $n$ variables $\mathbb{K}[\mathbf{x}] = \mathbb{K}[x_1, \dots, x_n]$} such that $\cl A = \kk[\bm x]/I$ is of finite dimension.
Conversely, any zero-dimensional ideal $I\subset \kkrng$ defines the Artinian algebra $\cl A= R/I$.

The algebraic set $\cl V_{{\okk}}(I)= \{\xi\in \okk^{n}\mid \forall p \in I, p(\bm \xi)=0\}$ is finite iff $I$ is zero dimensional, iff $\kkrng/I$ is Artinian, where $\okk$ is the algebraic closure of $\kk$ (see \cite{CLO}).

Let $r = \dim_\kk \cl A$, $\{\xi_{1}, \ldots, \xi_{r'}\}= \cl V_{{\okk}}(I)$ with $r'\le r$.
Let us assume that $\kk= \okk$ is algebraically closed and let $I = \cap_{i=1}^{r'} Q_i$  be a primary decomposition of the ideal $I$, where $Q_i$ is a primary ideal for the maximal ideal $\bm m_{\xi_i}$ defining $\xi_i$. Then $\cl A$ decomposes as 
$$
\cl A = \cl A_1 \oplus \cdots \oplus \cl A_{r'}
$$
where $\cl A_i = \bm u_i \cl A \sim R/Q_i$ is of dimension $\mu_i\le r$ and $\bm u_1, \ldots \bm u_{r'}$ is the family of orthogonal idempotents of maximal size, satisfying 
$\bm u_i^2= \bm u_i \textup{ for } i=1, \ldots, r', \quad \sum_{i=1}^{r'} \bm u_i = 1$.
The dimension $\dim_{\kk} \cl A_i$ is called the \emph{multiplicity} of $\xi_i$ (see \cite{Elkadi2007}).

Given $p\in R$ (or $p \in \cl A$ identifying $p\in R$ with its class in $\cl A=R/I$), we define the operator $M_{p}$ of multiplication by $p$ and its transposed $\dual{M_{p}}$ as 
$$
\begin{array}{rclcrcl}

M_{p}: \cl A  & \longrightarrow & \cl A  &  \qquad\qquad & \dual{M}_{p}:  \dual{\cl A} & \longrightarrow & \dual{\cl A}\\
       a  & \longmapsto & p\, a  &  & \Lambda & \longmapsto & p\star \Lambda \\
\end{array}
$$
where $\dual{\cl A}= \mathrm{Hom}_{\kk}(\cl A, \kk)$ is the dual of $\cl A$ and
$p \star \Lambda: q \in \cl A \longmapsto \Lambda(p\, q)\in \kk$.

We use the following classical theorem (see \cite{Auzinger1988,moller1995multivariate,Mourrain1998,Elkadi2007,Stetter2004}) for solving polynomial systems: 
\begin{theorem}
For any $p \in R=\kkrng$,  

\vspace{-5mm}
\begin{itemize}
    \item the eigenvalues of $M_{p}$ repeated with their multiplicities, are $\{\overbrace{p(\xi_{1}), \ldots, p(\xi_1)}^{\mu_1},$ \linebreak$\ldots, \overbrace{p(\xi_{r'}), \ldots, p(\xi_{r'})}^{\mu_{r'}}\}$, where $\mu_i$ is the multiplicity of the root $\xi_i$.
\item The common eigenvectors of all $(\dual{M}_{p})_{p\in R}$ are, up to a scalar, the evaluation functionals $\eval_{\xi_{i}}: q \in R \mapsto q(\xi_{i})$.
\end{itemize}
\end{theorem}

This allows us to recover the roots $\Xi= \set{\xi_1, \ldots, \xi_{r'}}$ by eigencomputation, using the operators of multiplication $M_{x_i}$ for $i=1, \dots, n$. In particular, when the roots are simple, we have the following (see e.g. \cite{Elkadi2007}): 
\begin{proposition}\label{prop:simple roots}If the roots $\set{\xi_1, \dots, \xi_r}= \cl V_{\okk}(I)$ are \emph{simple} (i.e. $\mu_i=1$),
\begin{itemize}
  \item  The operators $M_{x_i}$ for $i=1,\ldots, n$ have a joint diagonalization.
  \item Their common eigenvectors are, up to a scalar, the 
    \emph{interpolation polynomials}  $\bm u_{i}$ at the roots, satisfying $\bm u_i(\xi_j) = \delta_{i,j}$. They form a basis of orthogonal idempotents of $\cl A$ (i.e.  $\bm u_i^2\equiv \bm u_i$, $\bm u_i \bm u_j\equiv 0$ if $i\neq j$, $\sum_{i=1}^r  \bm u_i\equiv 1$).
\item The common eigenvectors of the transpose of the matrices of $(M_{x_i})_{i=1,\ldots, n}$ in a basis $B$ of $\cl A$ are, up to a scalar, the vectors
$B(\xi_j)$ representing the evaluation linear functionals $\eval_{\xi_{1}}, \ldots, \eval_{\xi_{r}}$ in the basis of 
$\dual{\cl A}= I^{\perp}$ dual to $B$. 
\end{itemize}
\end{proposition} 
When the roots are multiple, we have the 
following theorem:
\begin{proposition}\label{prop:mult roots}
There exists a basis of $\cl A$ in which, all the operators of multiplication $M_p$ for $p\in R$ decompose as $\mathrm{diag}(M_p^1, \ldots, M_p^{r'})$ where $M_{p}^{i}$ is the matrix of the operator
$$
M_{p}^{i}: a \in \cl A_i \longmapsto p \, a\in \cl A_i
$$
with $\mathrm{Trace}(M_{p}^{i}) = \mu_i\, p(\xi_i)$ and $\mu_i= \dim (\cl A_i)$ for $i=1, \ldots, r'$.
\end{proposition}
Such a basis is a basis adapted to the decomposition $\cl A = \oplus_{i=1}^{r'} \cl A_i$, i.e. of the form $\bm u_i \, b_{i,j}$ where $(b_{i,j})_{j}$ is a basis of $\cl A_i$ and $\bm u_i$ is the idempotent defining $\cl A_i$. See \cite{Corless1997,Elkadi2007} for more details. 

In order to analyse the multiplicity structure of the roots, we introduce the following notation. Let $\kk
[[y_1, \ldots, y_n]] =\kk [[\bm{y}]]$ be the ring of formal power series in the variables $y_1, \ldots, y_n$ with coefficients in the field
$\kk$. To simplify notation, we assume hereafter that \emph{$\kk$ is of characteristic $0$}.

There is a natural isomorphism between the ring of formal power series and the dual $\dual{R}= \mathrm{Hom}_{\kk} (R, \kk)$ of $R =\kk[x_1, \ldots,
x_n]$. It is given by the following pairing:
\begin{eqnarray*}
  \kk [[y_1, \ldots, y_n]] \times \kk [x_1, \ldots,  x_n] & \rightarrow & \kk\\
  (\bm{y}^{\alpha}, \bm{x}^{\beta}) & \mapsto
  & \left\langle \bm{y}^{\alpha} |  
  \bm{x}^{\beta} \right\rangle = \left\{ \begin{array}{ll}
    \alpha !  & \mathrm{if} \ \alpha = \beta\\
    0 & \mathrm{otherwise}.
  \end{array} \right.
\end{eqnarray*}
where $\alpha ! = \prod_{i=1}^{n} \alpha_i!$ for $\alpha = (\alpha_1, \ldots, \alpha_n) \in \NN^n$.
Namely, if $\Lambda \in \dual{R}= \mathrm{Hom}_{\kk} (R, \kk) $ is an element of the dual of $R=\kk
[\bm{x}]$, it can be represented by the series:
\begin{equation}
  \Lambda (\bm{y}) = \sum_{\alpha \in \NN^n} \Lambda
  (\bm{x}^{\alpha})  \frac{\bm{y}^{\alpha}}{\alpha ! }\in
  \kk [[\bm y]], \label{eq:dualfps}
\end{equation}
so that we have $\langle \Lambda (\bm{y}) \mid  \bm{x}^{\alpha} \rangle = \Lambda (\bm{x}^{\alpha})$.
The map $\Lambda \in R^{\ast} \mapsto \sum_{\alpha \in \NN^n} \Lambda
(\bm{x}^{\alpha})\,  \textcolor{black}{\frac{\bm{y}^{\alpha}} {\alpha!}} \in
\kk [[\bm{y}]]$ is an isomorphism \textcolor{black}{(see \cite{Mourrain2017} for more details)} and any series \ $\Lambda
(\bm{y}) = \sum_{\alpha \in \NN^n} \Lambda_{\alpha} \, 
\frac{\bm{y}^{\alpha}}{\alpha!} \in \kk [[\bm{y}]]$ can
be interpreted as a linear functional
\[ p = \sum_{\alpha \in A \subset \NN^n} p_{\alpha}
   \bm{x}^{\alpha} \in \kk [\bm{x}] \mapsto \langle \Lambda
   \mid p \rangle = \sum_{\alpha \in A \subset \NN^n} p_{\alpha}
   \Lambda_{\alpha}. \]
The coefficients $\Lambda_{\alpha} = \langle \Lambda \mid
\bm{x}^{\alpha} \rangle$ for $\alpha \in \NN^n$ are 
called the {\emph{pseudo-moments}} of $\Lambda$.
We identify the dual $\dual{R}=\mathrm{Hom}_{\kk} (\kk
[\bm{x}], \kk)$ with $\kk [[\bm{y}]]$.
Using this identification, the dual basis, denoted $(\bm y^{[\alpha]})_{\alpha \in \NN^{n}}$, of the monomial basis
$(\bm{x}^{\alpha})_{\alpha \in \NN^n}$ is $\left(
\frac{\bm{y}^{\alpha}}{\alpha!}\right)_{\alpha \in \NN^n}$: 
$\bm y^{[\alpha]} := \frac{\bm{y}^{\alpha}}{\alpha!}$.

From the definition, we verify that $\bm y^{\alpha} \in \dual{R}$ acts on the polynomials by derivation: $\forall p \in R, \scp{\bm y^{\alpha}\mid p}= \partial_{x_1}^{\alpha_1} \cdots \partial_{x_n}^{\alpha_n} (p)(0,\ldots, 0)$.

To illustrate the series representation, consider the evaluation at a point $\xi= (\xi_1, \ldots, \xi_n) \in \kk^n$: $\eval_{{\xi}} : p \in R  \longmapsto  p (\xi) \in \kk$. It is represented by the series:
  \begin{eqnarray*}
    \eval_{{\xi}} (\bm{y}) & = &  
 \sum_{\alpha \in \NN^n}
    \xi^{\alpha}  \frac{\bms{y}^{\alpha}}{\alpha !}
    = 
    e^{\xi_1 \bms y_1 + \cdots + \xi_n \bms y_n} = e^{\langle \xi, \bms y \rangle},
\end{eqnarray*}
that is, an exponential series in the basis $(\bms y^{\alpha})_{\alpha\in \NN^n}$.

\begin{definition}[{Polynomial-Exponential series}]
$$ 
\PolExp(\bm y)= \left\{ \Lambda ( \bm y) = \sum_{i = 1}^r \omega_{i} (\bm y) \,
\eval_{\xi_i} ( \bm y) \mid \omega_i(\bm y)\in \kk[\bm y], \xi_{i}\in \kk^{n} \right\}
\vspace{-0.2cm}
$$
where $\eval_{\xi_i} ( \bm y)$ \emph{is} the series representing the evaluation $\eval_{\xi_i}: p\in
R \mapsto p(\xi_i)$ at $\xi_i$.
\end{definition}


For an ideal $I \subset R$, we denote by $I^{\bot}
\subset \dual{R}$ the space of linear forms $\Lambda \in
\dual{R}$, such that $\forall p \in I$, $\langle \Lambda
\mid p \rangle = 0$. Similarly, for a vector space $D \subset \kk
[[\bm{y}]]$, we denoted by $D^{\bot} \subset \kk [\bm{x}]$
the space of polynomials $p \in \kk [\bm{x}]$, such that
$\forall \Lambda \in D$, $\langle \Lambda \mid p \rangle = 0$. 

The dual space $\dual{R}$ has a natural structure of $R$-module, defined as follows: $\forall \Lambda \in \dual{R}, \forall p, q  \in R$,
\begin{eqnarray*}
  \langle p \star \Lambda   \mid q \rangle & = & \langle \Lambda 
   \mid p  q \rangle.
\end{eqnarray*}

We verify that $x_i$ acts on the series in $\bm y$ by derivation: $\forall \Lambda \in \kk[[\bm y]], x_i \star \Lambda= \frac{\partial}{\partial_{y_i}}(\Lambda)$.
\begin{definition}[{Inverse system}]
For $\bms \Lambda \subset \dual{R}$, the \emph{inverse system} spanned by $\bms \Lambda$ is 
$$
\invsyst{\bms \Lambda}= \vspan{p \star \Lambda; p\in R, \Lambda  \in \bms \Lambda}
$$
\end{definition}
As $x_i$ acts as a derivation on the series $\in \kk[\bm y]]$,
the inverse system generated by $\bm \Lambda$ is represented
by the vector space spanned by all possible derivations $\partial_{\bm y}^\alpha(\Lambda(\bm y))$ for all $\alpha\in \NN^n$, $\Lambda \in \bms \Lambda$.
If $\Lambda(\bm y)\in \PolExp(\bm y)$, we verify that $\invsyst{\Lambda}$ is finite dimensional. 
See \cite{Mourrain2017}.

\begin{theorem}Assume that $\kk=\okk$  is algebraically closed and let $\cl A= \cl R/I$ be an Artinian algebra with 
$\cl V(I)=\{\xi_{1},\ldots,\xi_{r'}\}$ and 
$I=Q_{1}\cap \cdots \cap Q_{r'}$ where $Q_{i}$ are the $\bm m_{\xi_{i}}$-primary components of $I$. Then, 
$$ 
\dual{\Ac}  = \oplus_{i=1}^{r'}\,\Dc_{i} \, \eval_{\xi_{i}}(\bm y) \subset \PolExp(\bm y)
$$
\begin{itemize}
 \item $\Dc_{i} = \dual{\cl A_i}
 = \invsyst{\omega_{i,1}(\bm y), \ldots, \omega_{i,l_{i}}(\bm y)}$ with $\omega_{i,j}(\bm y)\in \kk[\bm y]$.
 \item $\dim_{\kk} (\Dc_{i})= \mu_{i}$ multiplicity of $\xi_{i}$.
\end{itemize}
where $\cl D_{i}$ is the \emph{inverse system} generated by $\omega_{i,1}(\bm y), \ldots, \omega_{i,l_{i}}(\bm y)\in \kk[\bm y]$
$$ 
\invsyst{\omega_{i,1}(\bm y), \ldots, \omega_{i,l_{i}}(\bm y)}=\langle \partial_{\bm y}^{\alpha}
(\omega_{i,j}) , \alpha\in \NN^{n}, j=1, \ldots, l_{i} \rangle
$$ 
\end{theorem}

For symbolic-numeric algorithms computing the inverse system $\cl D_i = \dual{\cl A_i}= Q_i^{\perp}$ at $\xi_i$ from the generators of $I$, see \cite{Mourrain1998,hauenstein:hal-01250388,mantzaflaris:hal-03079910,mantzaflaris:hal-02478768} and Section \ref{sec:mult roots}.

\section{Computing effectively these structures} \label{sec:examples}

From a computational point of view, we handle the algebraic structure of an Artinian algebra, via so-called \emph{Truncated Normal Forms}:
\begin{definition}[Truncated Normal Form] \label{def:normalform}
Let $W \subset R$ and $V$  be a $\kk$-vector space. A \textup{Truncated Normal Form (TNF)} for the ideal $I$ from $W$ to $V$ is a linear map $\bm N: W \rightarrow V$ such that the sequence
$$
\begin{array}{rcl}
    0 \longrightarrow  K \longrightarrow  W & \stackrel{\bm N}{\longrightarrow} & V \longrightarrow 0\\
\end{array}
$$
is exact, $I= (K)$ is the ideal of $R$ generated by $K:= \ker \bm N$,  $ I \cap W = K$ and $I + W = R$.
\end{definition}
For other characterisations of TNF, see e.g. \cite{Telen2018,Mourrain2021}. 
By definition, if $N$ is a TNF, then it induces an isomorphism between $\Ac = R/(\ker \bm N)$ and $V$.
For any set $B\subset R$, let $B^+= B \cup x_1\cdot B \cup \cdots \cup x_n \cdot B$.

The multiplicative structure of an Artinian algebra $\cl A = R/I$ can be recovered using the following result:
\begin{proposition}\label{prop:tnf:mult}
Let $\bm N:W\rightarrow V$ be a TNF such that there exists $B\subset W$ with $B^{+} \subset W$ and $\bm N_{|B}: B\rightarrow V$ bijective. Let $I = (\ker \bm N)$. Then
$$ 
\begin{array}{rcl}
\bm  M_{i} : B & \rightarrow & {B}\\
    b & \mapsto & (\bm N_{|B})^{-1} \circ \bm N(x_{i} b).
\end{array}
$$
is the operator of multiplication by $x_{i}$ in the quotient $R/I$.
\end{proposition}




Hereafter is the general form of the eigensolver algorithm, that is implemented in \AlgebraicSolvers:
\begin{algorithm}[H]
\caption{\textsc{eigensolve}\label{algo:solve}}
\begin{algorithmic}[1]

\Function{eigensolve}{$\bm P$}
    \State \textbf{Input}: $\bm P=(P_1, \ldots, P_m)\in R^m$
    \State Compute a Truncated Normal Form $N : W\longrightarrow V$ for $I=(P_1, \ldots, P_m)$; 
    \State Extract a basis $B\subset W$ such that $N_{|B}$ is bijective and $B^+\subset W$;
    \State Build multiplication matrices $M = \{M_{x_1},\dots,M_{x_n}\}$ in the basis $B$ from $N$;
    \State Compute the eigenvalues $\bms \lambda$ of a random combination $M_{\mathrm{rnd}}$ of $M_1, \ldots, M_n$ and their multiplicities;
    \If { the eigenvalues are simple}
        \State Compute the points $\Xi$ by joint diagonalisation of $M$. 
        \Else
        \State Compute the points $\Xi$ and their multiplicities $\bms \mu$ by a joint Schur factorization of $M$.
        \EndIf
       \State \textbf{Output}: the roots $\Xi$ and their multiplicities $\bms \mu$.  
\EndFunction
\end{algorithmic}
\end{algorithm}

Determining the multiplicity of the eigenvalues $\bms \lambda$ (step 6) is performed by a clustering function that uses a threshold $\epsilon$ (\texttt{{1.e-5}} by default) to identify points in the same cluster.

The joint diagonalization (step 8) is implemented as an iterative process, optimizing the sum of the square off-diagonal norm of the matrices $E^{-1}\,M_i E$ with respect to the (unknown) eigenvector matrix $E$, starting from the eigenvector matrix of $M_{\mathrm{rnd}}$ in Algorithm \ref{algo:solve} (see \cite{vanderhoeven:hal-01579079}).

The joint Schur factorization (step 10) applies the Schur Factorization of $M_{\mathrm{rnd}}$ to $M_1, \ldots, M_n$ and uses the result of the clustering function to deduce the operators of multiplication by the variables $x_i$ in the local algebras associated to the roots.
Their coordinates are recovered from the trace of these local multiplications (see Proposition \ref{prop:mult roots}). Their multiplicities $\bms \mu$ are the sizes of the clusters.

Hereafter, we illustrate these techniques and their implementation in the package \AlgebraicSolvers\footnote{See also its documentation at \href{https://algebraicgeometricmodeling.github.io/AlgebraicSolvers.jl/}{https://algebraicgeometricmodeling.github.io/AlgebraicSolvers.jl/}.} on few examples.

 
\subsection{Solving polynomial systems}\label{sec:mult roots}

In this section, we illustrate the computation of roots from eigen computations, in the presence of multiple roots. We consider the following system of two equations in two variables \(x_1, x_2\):

\begin{lstlisting}
using DynamicPolynomials, LinearAlgebra, AlgebraicSolvers, Groebner
x = (@polyvar x[1:2])[1]
P = [x[1]^2 + x[1] - x[2], x[2]^2 + x[1] - x[2]]
\end{lstlisting}

\begin{lstlisting}[language={},xleftmargin=-\fboxsep,xrightmargin=-\fboxsep,backgroundcolor=\color{white},frame=single]
 -x_2 + x_1 + x_1^2
 -x_2 + x_1 + x_2^2
\end{lstlisting}

We compute a TNF via Grobner basis computation with exact coefficients (using the package \texttt{Groebner.jl} \cite{demin2023groebner}):

\begin{lstlisting}
 GB = Grobner(
    Groebner.DegRevLex,      # function of the variables defining the monomial ordering
    Groebner.groebner,       # function for computing Grobner basis
    Groebner.normalform,     # function that computes the normal form of a polynomial
    Groebner.quotient_basis  # function that computes a basis of the quotient algebra
)
N, L = tnf(P, GB); round.(N, digits=6)
\end{lstlisting}

\begin{lstlisting}[language={},xleftmargin=-\fboxsep,xrightmargin=-\fboxsep,backgroundcolor=\color{white},frame=single]
4×8 Matrix{Float64}:
 1.0  0.0  0.0  0.0   0.0   0.0   0.0   0.0
 0.0  1.0  0.0  0.0   1.0   1.0  -1.0   1.0
 0.0  0.0  1.0  0.0  -1.0  -1.0   1.0  -1.0
 0.0  0.0  0.0  1.0   0.0   0.0   1.0  -1.0
\end{lstlisting}

Let \(B=[1, x_1, x_2, x_1 x_2]\) be the first \(4\) elements of \(L\). Then we verify that \(L= B^+\):

\begin{lstlisting}[language={},xleftmargin=-\fboxsep,xrightmargin=-\fboxsep,backgroundcolor=\color{white},frame=single]
(1, x_2, x_1, x_1*x_2, x_2^2, x_1^2, x_1*x_2^2, x_1^2*x_2)
\end{lstlisting}

The matrix \texttt{N} represents the TNF map \(N: W \longrightarrow \mathbb{R}^4\) where \(W = \langle L \rangle\). It is computed as follows:

\begin{itemize}
\item first we compute a Grobner basis \texttt{G}  of \texttt{P}, using the function \texttt{Groebner.groebner} with the monomial ordering \texttt{Groebner.DegRevLex(variables(P))}. The coefficients of the polynomials in this computation are rational \(\in \mathbb{Q}\);

\item then we compute a basis \texttt{B} of the quotient by the ideal \(I=(P_1, P_2)\) using \linebreak\texttt{Groebner.quotient\_basis};

\item Finally we compute the reduction of the monomials in \(L\) by \texttt{G} (using \texttt{Groebner.normalform}) and decompose the remainder in the basis \texttt{B} using \texttt{Float64} numbers. This gives the columns of \texttt{N}.

\end{itemize}

We compute the multiplication matrices by the variables \(x_1, x_2\) in the basis \(B\), using the TNF \texttt{N} indexed by the monomials \(L\):

\begin{lstlisting}
M = mult_matrices(N, L , collect(1:4), x); round.(M[1], digits=6)
\end{lstlisting}

\begin{lstlisting}[language={},xleftmargin=-\fboxsep,xrightmargin=-\fboxsep,backgroundcolor=\color{white},frame=single]
4×4 Matrix{Float64}:
 0.0  0.0   0.0   0.0
 0.0  0.0   1.0   1.0
 1.0  0.0  -1.0  -1.0
 0.0  1.0   0.0  -1.0
\end{lstlisting}

To obtain the roots and their multiplicities, we perform a joint Schur factorization:

\begin{lstlisting}
Xi, ms, Z = schur_dcp(M); round.(Xi, digits=6)
\end{lstlisting}

\begin{lstlisting}[language={},xleftmargin=-\fboxsep,xrightmargin=-\fboxsep,backgroundcolor=\color{white},frame=single]
2×2 Matrix{ComplexF64}:
 -0.0+0.0im  -2.0+0.0im
 -0.0+0.0im   2.0+0.0im
\end{lstlisting}

The solutions\brev{}, which can be obtained directly using the function \texttt{solve(P,GB)},\erev{} are the columns of \brev{}the matrix\erev{} \texttt{Xi}. The corresponding multiplicities are \texttt{length.(ms)}:

\begin{lstlisting}[language={},xleftmargin=-\fboxsep,xrightmargin=-\fboxsep,backgroundcolor=\color{white},frame=single]
 3  1
\end{lstlisting}
and the Schur factor is \texttt{Z}. 

Instead of using Grobner basis computation to get a TNF, we could also have used a resultant matrix construction 
\brev
(hereafter Macaulay construction):
\begin{lstlisting}
R, L = res_matrix(P, Macaulay()); round.(R, digits=6)
\end{lstlisting}
\begin{lstlisting}[language={},xleftmargin=-\fboxsep,xrightmargin=-\fboxsep,backgroundcolor=\color{white},frame=single]
6×10 SparseArrays.SparseMatrixCSC{Float64, Int64} with 18 stored entries:
  .   -1.0  1.0    .     .   1.0   .    .    .    . 
  .     .    .   -1.0   1.0   .    .    .   1.0   . 
  .     .    .     .   -1.0  1.0   .    .    .   1.0
  .   -1.0  1.0   1.0    .    .    .    .    .    . 
  .     .    .   -1.0   1.0   .   1.0   .    .    . 
  .     .    .     .   -1.0  1.0   .   1.0   .    . 
\end{lstlisting}

The rows represent the monomial multiples of degree \(\le 3\) of the equations \texttt{P[i]}.  It is a matrix of size \(6\times 10\), which columns are indexed by the monomials:
\begin{lstlisting}[language={},xleftmargin=-\fboxsep,xrightmargin=-\fboxsep,backgroundcolor=\color{white},frame=single]
(1, x_2, x_1, x_2^2, x_1*x_2, x_1^2, x_2^3, x_1*x_2^2, x_1^2*x_2, x_1^3)
\end{lstlisting}
The corresponding TNF, using the function \texttt{tnf(P,Macaulay())}, is the kernel of \texttt{R}:
\begin{lstlisting}[language={},xleftmargin=-\fboxsep,xrightmargin=-\fboxsep,backgroundcolor=\color{white},frame=single]
4×10 Matrix{Float64}:
 1.0  0.0  0.0  -0.0  0.0  -0.0  -0.0  -0.0  -0.0  -0.0
 0.0  0.0  0.0  -0.0  1.0  -0.0  -1.0   1.0  -1.0   1.0
 0.0  0.0  1.0  -1.0  0.0  -1.0  -1.0   1.0  -1.0   1.0
 0.0  1.0  0.0   1.0  0.0   1.0   1.0  -1.0   1.0  -1.0
\end{lstlisting}

Using this TNF yields the same basis $B$ and a numerical accuracy on the roots  of the same order as the one obtained via Grobner basis computation.
\erev

We analyze now the inverse system of the multiple point \(\xi_0=[0,0]\):

\begin{lstlisting}
D, B0 = invsys(P,[0,0]); D
\end{lstlisting}

\begin{lstlisting}[language={},xleftmargin=-\fboxsep,xrightmargin=-\fboxsep,backgroundcolor=\color{white},frame=single]
 1
 dx_1 + dx_2
 dx_1^2 + dx_2 + dx_1*dx_2 + dx_2^2
\end{lstlisting}

These series represent a basis of the inverse system of \(I=(P_1, P_2)\) at the point \(\xi_0\). They are described as series (\texttt{AlgebraicSolvers.Series}), in terms the dual basis \(dx_1^{\alpha_1}dx_2^{\alpha_2}\) of the monomial basis \(\mathbf{x}^{\alpha}=x_1^{\alpha_1}x_2^{\alpha_2}\).

A basis of the local algebra \(\mathcal A_{\xi_0}\) is \texttt{B0 = $(1, x_1, x_1^2)$}. We verify that \texttt{B0} is dual to \texttt{D}. 
This inverse system is generated by \texttt{D[3]}, since the following series span the same space of series as \texttt{D}:

\begin{lstlisting}
x[1]^2*D[3], x[2]^2*D[3], x[1]*x[2]*D[3], x[1]*D[3], x[2]*D[3], D[3]
\end{lstlisting}

\begin{lstlisting}[language={},xleftmargin=-\fboxsep,xrightmargin=-\fboxsep,backgroundcolor=\color{white},frame=single]
(1, 1, 1, dx_1 + dx_2, 1 + dx_1 + dx_2, dx_1^2 + dx_2 + dx_1*dx_2 + dx_2^2)
\end{lstlisting}

\subsection{Decomposing symmetric tensors}\label{sec:tensor dec}

The \emph{Generalized Additive Decomposition} (GAD) of a symmetric tensor, or equivalently a homogeneous polynomial  (or form) $F$ of degree $d$ in the variables 
$\bm x = (x_0, \ldots, x_n)$ is a decomposition of the form
\begin{equation}\label{eq:gad}
F = \sum_{i=1}^{r} \omega_i(\bm x)\, (\xi_i \cdot \bm x)^{d-k_i}
\end{equation}
where $\omega_i(\bm x)$ is a homogeneous polynomial in the variables $\bm x$ of degree $k_i$ with $0 \leq k_i \leq d$, and $\xi_i \in \CC^{n+1}$. When all the degrees $k_i$ vanish, such a decomposition is called a \emph{Waring decomposition} and the minimal $r$ in a Waring decomposition is called the \emph{rank} of $F$. See \cite{barrilli:hal-05339397} for more details on the GAD algorithm.

\begin{lstlisting}
using DynamicPolynomials, LinearAlgebra, AlgebraicSolvers, TensorDec
X  = (@polyvar x[0:2])[1]; x = X[2:end]
F = (X[1]-2*X[2]+ 2*X[3])^5 + (X[1]*X[3]+ X[2]^2+2*X[2]*X[3]+ X[3]^2)*X[1]^3
\end{lstlisting}
{\small
$$ 
\begin{array}{l}
32x_{2}^{5} - 160x_{1}x_{2}^{4} + 320x_{1}^{2}x_{2}^{3} - 320x_{1}^{3}x_{2}^{2} + 160x_{1}^{4}x_{2} - 32x_{1}^{5} + 80x_{0}x_{2}^{4} - 320x_{0}x_{1}x_{2}^{3} + 480x_{0}x_{1}^{2}x_{2}^{2} \\\quad - 320x_{0}x_{1}^{3}x_{2}  + 80x_{0}x_{1}^{4} + 80x_{0}^{2}x_{2}^{3} - 240x_{0}^{2}x_{1}x_{2}^{2} + 240x_{0}^{2}x_{1}^{2}x_{2} - 80x_{0}^{2}x_{1}^{3} + 41x_{0}^{3}x_{2}^{2} - 78x_{0}^{3}x_{1}x_{2} + 41x_{0}^{3}x_{1}^{2} 
\\\quad + 11x_{0}^{4}x_{2} - 10x_{0}^{4}x_{1} + x_{0}^{5} 
\end{array}
$$
}

We compute the series associated with the tensor \texttt{F} in degree $d$ by apolar duality:

\begin{lstlisting}
sigma = apolar_dual(F)
\end{lstlisting}

\begin{lstlisting}[language={},xleftmargin=-\fboxsep,xrightmargin=-\fboxsep,backgroundcolor=\color{white},frame=single]
32dx_2^5 - 32dx_1*dx_2^4 + 32dx_1^2dx_2^3 - 32dx_1^3dx_2^2 + 32dx_1^4dx_2 - 32dx_1^5 + 16dx_0*dx_2^4 - 16dx_0*dx_1*dx_2^3 + 16dx_0*dx_1^2*dx_2^2 - 16dx_0*dx_1^3*dx_2 + 16dx_0*dx_1^4 + 8dx_0^2dx_2^3 - 8dx_0^2dx_1*dx_2^2 + 8dx_0^2dx_1^2dx_2 - 8dx_0^2dx_1^3 + 410dx_0^3dx_2^2 - 390dx_0^3dx_1*dx_2 + 410dx_0^3dx_1^2 + 11//5dx_0^4dx_2 - 2dx_0^4dx_1 + dx_0^5
\end{lstlisting}
\brev{}
where $dx^{\alpha}$ are the linear functionals  $\bm y^{[\alpha]}= \frac{\bm y^{\alpha}}{\alpha!}$ dual to the monomial basis $\bm x^{\alpha}$.\erev{}
We deduce the Hankel (or Catalecticant) matrix in degree $2,3$. From this, by Theorem 3.9 of \cite{mourrain:hal-05135952} we recover the TNF \texttt{N} by extracting the submatrix corresponding to the rows indexed by the monomial basis \(B= (1, x_1, x_2, x_1x_2)\), which, in the homogenized setting, corresponds to \((x_0^2, x_0x_1, x_0x_2, x_1x_2)\).

\begin{lstlisting}
L2 = reverse(monomials(X,2)); L3 = reverse(monomials(X,3))
Hk = hankel(sigma,L2,L3)
I = [1,2,3,5]; N = Hk[I,:]
\end{lstlisting}

\begin{lstlisting}[language={},xleftmargin=-\fboxsep,xrightmargin=-\fboxsep,backgroundcolor=\color{white},frame=single]
4×10 Matrix{Rational{Int64}}:
    1       -2      11//5   41//10  -39//10  41//10   -8    8   -8    8
   -2      41//10  -39//10   -8        8      -8      16  -16   16  -16
  11//5   -39//10   41//10    8       -8       8     -16   16  -16   16
 -39//10     8       -8     -16       16     -16      32  -32   32  -32
\end{lstlisting}

At this point we can compute the multiplication matrices by the variables \(x_1, x_2\), while substituting $x_0$ by $1$:

\begin{lstlisting}
M = mult_matrices(N, subs.(L3, X[1] => 1), I, x)
\end{lstlisting}

\begin{lstlisting}[language={},xleftmargin=-\fboxsep,xrightmargin=-\fboxsep,backgroundcolor=\color{white},frame=single]
2-element Vector{Matrix{Rational{Int64}}}:
 [0 0 0 0; 1 -2//3 0 -2//3; 0 2//3 0 2//3; 0 -1//3 1 -4//3]
 [0 0 0 0; 0 0 -2//3 2//3; 1 0 2//3 -2//3; 0 1 -1//3 4//3]
\end{lstlisting}

We recover the points and their multiplicities by performing a joint Schur factorization.

\begin{lstlisting}
Xi, ms = schur_dcp(M, 1.e-3); Xi = vcat(ones(1, size(Xi,2)), Xi)
\end{lstlisting}

\begin{lstlisting}[language={},xleftmargin=-\fboxsep,xrightmargin=-\fboxsep,backgroundcolor=\color{white},frame=single]
3×2 Matrix{ComplexF64}:
  1.0+0.0im  1.0+0.0im
 -2.0+0.0im  0.0+0.0im
  2.0+0.0im  0.0+0.0im
\end{lstlisting}

We may assume, without loss of generality, that the points do not lie in the hyperplane $\{x_0
=0\}$. Indeed, over an infinite field, a generic linear change of coordinates ensures that the first coordinate of each point $\xi_i$ is nonzero. After rescaling, we may further assume that these first coordinates are all equal to 1. We have added a first row of 1's to obtain the corresponding projective points.
The linear forms in the GAD decomposition \eqref{eq:gad} are \(L_i  = (\xi_i \cdot \bm x)\) where $\xi_i=$\texttt{Xi[:,i]} is the i$^{\mathrm{th}}$ column of \texttt{Xi}.

We are left with computing \(k\), the vector of degrees of the homogeneous polynomials \(\omega_i\){\textquotesingle}s. We compute it from the nil-indices of the multiplication operators in the local algebras $\cl A_{\xi_i}$ (see \cite{barrilli:hal-05339397}) and get 

\begin{lstlisting}[language={},xleftmargin=-\fboxsep,xrightmargin=-\fboxsep,backgroundcolor=\color{white},frame=single]
(0, 2)
\end{lstlisting}

To complete the decomposition, we compute the coefficients of the  \(\omega_i\){\textquotesingle}s, by solving a Vandermonde-like linear system in the least-square sense (see \cite{barrilli:hal-05339397}) and get \texttt{W}:

\begin{lstlisting}[language={},xleftmargin=-\fboxsep,xrightmargin=-\fboxsep,backgroundcolor=\color{white},frame=single]
((1.0 + 0.0im), x_2^2 + (2.0 + 0.0im)x_1*x_2 + x_1^2 + x_0*x_2)
\end{lstlisting}

These weights represent the generators of the inverse systems of the local algebras associated with $F$. We can check that the full Artinian algebra associated with $F$ is isomorphic to the one defined in Section \ref{sec:mult roots}, by comparing the points and the generators of their inverse systems.

We verify the accuracy of the reconstruction by forming the reconstructed tensor from the linear forms $L$ and the weights \texttt{W} of degree \(k_i\).

\begin{lstlisting}
T = sum(W[i]*L[i]^(maxdegree(F)-k[i]) for i in 1:length(W))
e = apolar_norm(F-T)
\end{lstlisting}

\begin{lstlisting}[language={},xleftmargin=-\fboxsep,xrightmargin=-\fboxsep,backgroundcolor=\color{white},frame=single]
9.342121376789858e-14
\end{lstlisting}

The reconstruction error is of order \(10^{-13}\), demonstrating the accuracy of the reconstruction.


\textbf{Acknowlegdment.}
This work has been supported by European Union’s HORIZON–MSCA-2023-DN-JD programme under the Horizon Europe (HORIZON) Marie Skłodowska-Curie Actions, grant agreement 101120296 (TENORS).
\end{document}